\documentstyle[aps,12pt]{revtex}

\begin{document}
\author{Z. Yusof}
\title{Modeling of High-Tc Tunneling Conductance.}
\date{9-20-1997}

\begin{center}
{\LARGE Modeling of Tunneling Spectroscopy in High-T}$_{c}${\LARGE \
Superconductors, Incorporating Band Structure, Gap Symmetry, Group Velocity,
and Tunneling Directionality.\bigskip }

\bigskip {\large Z. Yusof, J.F. Zasadzinski, L. Coffey\bigskip }

{\it Science and Technology Center for Superconductivity,}

{\it Illinois Institute of Technology, Chicago Illinois 60616\bigskip
\bigskip }

\medskip {\large ABSTRACT\bigskip \negthinspace }
\end{center}

A theoretical model for tunneling spectroscopy employing tight-binding band
structure, $d_{x^{2}-y^{2}}$ gap symmetry, group velocity, and tunneling
directionality is studied. This is done to investigate if the model can
exhibit the same wide range of characteristics observed in tunneling
experiments on high-T$_{c}$ superconductors. A band structure specific to
optimally-doped Bi$_{2}$Sr$_{2}$CaCu$_{2}$O$_{8}$ (BSCCO) is used to
calculate the tunneling density of states for a direct comparison to
experimental tunneling conductance. A robust feature of the model is an
asymmetric, decreasing conductance background, which is in agreement with
experiment for BSCCO. The model also produces generally good agreement with
the tunneling data especially in the gap region. In particular, the
experimentally observed asymmetric conductance peaks can be understood with
this model as a direct consequence of the $d_{x^{2}-y^{2}}$ gap symmetry.
Dip features observed at $|eV|\sim 2\Delta $ in the experimental data are
not found for any range of parameters in this model, indicating that these
features are caused by other physical mechanisms such as strong coupling
effects.\bigskip

\newpage

\section{INTRODUCTION}

Tunneling measurements on high-T$_{c}$ superconductors (HTS) have revealed a
rich variety of properties and characteristics. While tunneling
spectroscopies on conventional superconductors can directly reveal the
density of states (DOS) of the superconductor\cite{Wolf}, the same
measurements in HTS are not as easily interpreted. This is compounded by the
lack of consensus on a single theory for the mechanism causing
superconductivity in these materials. There are however, a set of
characteristics that are consistently observed in a wide variety of
tunneling measurements. These include : (i) variable sub-gap shape, ranging
from sharp, cusp-like\cite{SPIE}, to a flat, BCS-like feature\cite{Chen}\cite
{Yannick}; (ii) symmetric\cite{SPIE}\cite{Renner} and asymmetric conductance
peaks\cite{Chen}\cite{Yannick}\cite{Hancotte}; (iii) variable conductance
shape outside the gap region (background), which can be flat, decreasing, or
increasing values at high bias voltages\cite{SPIE}\cite{Yannick}\cite{Renner}%
\cite{Huang}\cite{Hasegawa} with varying degrees of asymmetry; (iv) dip
features at $|eV|\sim 2\Delta $\cite{Yannick}\cite{Renner}\cite{Hancotte}.

Sometimes, a range of varying features can be seen using the same technique
and on the same HTS sample. This is illustrated by the two Bi$_{2}$Sr$_{2}$%
CaCu$_{2}$O$_{8}$ (BSCCO) tunneling conductance curves shown in Fig. 1\cite
{Yannick}. These are superconductor-insulator-normal metal (SIN) conductance
data obtained using the point-contact tunneling (PCT) method on the same
optimally-doped BSCCO crystal. Fig. 1a shows a conductance curve with sharp,
cusp-like sub-gap feature at zero-bias, and approximately symmetric
conductance peaks as would be expected from a momentum-averaged $%
d_{x^{2}-y^{2}}$ (d-wave) DOS. On the other hand, the data shown in Fig. 1b
has a much flatter, BCS-like sub-gap conductance, and a strongly asymmetric
conductance peaks. Both conductance curves show distinctive dip features at $%
eV\sim -2\Delta $ and a decreasing, asymmetric background conductance
outside the gap. The two curves of Fig. 1 display characteristics that are
also seen in other tunneling data. Vacuum tunnel junction on BSCCO from
scanning tunneling microscopy (STM)\cite{Yannick}\cite{Renner} display gap
features in between the two extremes of Figs. 1a and 1b and more symmetric
dip features. It is apparent that data collected from identical samples
using the same technique can yield a variety of tunneling conductance curves
which may at first seem contradictory. Any model that tries to explain
superconductivity in HTS must be able to consistently account for these
variable features observed in tunneling measurements.

It is found experimentally that the sharpest gap features are obtained when
the high-bias background is weakly decreasing as shown in Fig. 1\cite{SPIE}%
\cite{Yannick}\cite{Renner}. A quantitative measure is the conductance peak
height to background (PHB) ratio which is $>2$ for both spectra of Fig. 1.
When the background conductance is linearly increasing ($\propto |V|$), the
PHB ratio is invariably much less than 2\cite{SPIE}. This suggests that the
linearly increasing background is arising from an additional conductance
channel such as inelastic tunneling\cite{Kouznetsov}\cite{Kirtley}.

Another indication that the decreasing background is an intrinsic effect of
elastic tunneling in HTS is that it cannot be explained by any extrinsic
processes such as junction heating, which might be a concern with local
probes such as PCT or STM. PCT junctions on Nb\cite{Huang2} showed no
evidence of heating whatsoever. Also, the heating power is $V^{2}/R_{J}$ and
considering a fixed voltage of say $V=200$ meV, the effects of heating
should scale with junction resistance $R_{J}$. The PCT junction of Fig. 1a
has $R_{J}\sim 2$ k$\Omega $ whereas STM junctions are typically 10$^{9}$ $%
\Omega $\cite{Yannick}\cite{Renner}. Yet both PCT and STM junctions show a
similar decreasing background. It is unlikely that junction area alone would
account for the factor of 10$^{6}$ difference in PCT and STM junctions
(leading to fixed power/area) and therefore heating is ruled out as the
explanation of decreasing background.

A standard technique in analyzing the tunneling conductance data in HTS is
to use a smeared BCS function\cite{Dynes} in which a scattering rate
parameter, $\Gamma $, is used to account for any broadening of the gap
region in the DOS. This method of analysis suffers from the deficiency that
it does not adequately treat the problem of a gap with d-wave symmetry, and
it cannot explain the asymmetry observed in tunneling conductances. In
addition, it also requires that the comparison be made with normalized
tunneling conductance data. This is done by dividing the experimentally
measured conductance by its ``normal state'' value which is usually obtained
by extrapolating the high bias background conductance down to zero bias.
However, since HTS tunneling conductance can exhibit varied and complex
background shape, this procedure may ``filter out'' too much information
from of the conductance data. This paper will deal with the problem of
analyzing HTS conductance data from a different perspective. Rather than
manipulating or normalizing the conductance data in order to fit the smeared
BCS model, the experimental conductance data are left alone, and the model
is adapted to match the experimental data. The model that will be presented
has three general features: (i) a realistic band structure for the $Cu-O_{2}$
plane in HTS; (ii) a tunneling matrix element incorporating directionality
and group velocity\cite{Harrison}; (iii) gap symmetry.

We will apply this model to optimally-doped BSCCO by using a band structure
specific to this HTS obtained from angle-resolved photoemission spectroscopy
(ARPES)\cite{Norman}. Since there is an emerging consensus that the pairing
interaction in HTS has a d-wave symmetry, the DOS calculated with this model
will use this gap symmetry. This result will then be compared to the two
different BSCCO data shown earlier. It is emphasized that this comparison is
done with the {\it unnormalized} experimental data.

One of the early theoretical model calculations for the normal state and
superconducting DOS of HTS was done by Fedro {\it et al}.\cite{Fedro} using
the tight-binding band structure $\xi _{{\bf k}}$ including the next-nearest
neighbor hopping ({\it t}-{\it t}' model) to describe the electrons in the $%
Cu-O_{2}$ planes \cite{tetragonal}\mathstrut \medskip

\[
\xi _{{\bf k}}=-2t[\cos (k_{x}a)+\cos (k_{y}a)]+4t^{\prime }\cos
(k_{x}a)\cos (k_{y}a)-\mu 
\]
\medskip where {\it t} and {\it t}' are nearest-neighbor and next
nearest-neighbor hopping energies respectively, and $\mu $ is the chemical
potential. Both $\Delta =\Delta _{o}$ (s-wave) and d-wave gap symmetries
were considered. The DOS with the s-wave gap symmetry showed the expected
BCS-like gap shape with flat sub-gap feature, while the DOS with the d-wave
gap symmetry showed a V-shaped, cusp-like feature at zero-bias. An
interesting feature of this DOS calculation is the presence of as many as
two singularities in the normal state. When $t^{\prime }=\mu =0$, there
exists a van Hove singularity (VHS) in the center of the band (Fig. 1a of
Ref. 15). This is due to the saddle-point in {\it k}-space near ($\pi $,0)
in the band structure. For the hole-doped situation ($\mu <0$), this
singularity exists at an energy higher than the gap. When the {\it t}' term
is considered, the VHS shifts to below the energy gap region of the DOS.
There is also a second singularity at the lower edge of the energy band
(Fig. 1c and 1d of Ref. 15). This additional singularity is caused by the
extra flattening-out of the energy band at ${\bf k}=(0,0)$. However, other
than the quasiparticle peaks at $eV=\pm \Delta $, no HTS tunneling
measurement has displayed any other peaks in the DOS as prominent and
distinctive as those shown in the {\it t}-{\it t}' model DOS calculation of
Ref. 15. Fig. 1a, which is a typical curve for BSCCO, shows no direct
evidence of a VHS either inside or outside the gap.

It was proposed by Harrison\cite{Harrison} that tunneling measurements are
insensitive to band structure effects. Although the tunneling matrix element
is usually considered to be constant in the conventional treatment of
tunneling, a more careful consideration of the matrix element reveals a need
for a factor $\nabla \xi _{{\bf k}}$ (which is proportional to the
particles' group velocity {\it v}$_{g}$ and thus will be called as such) and
directionality in the tunneling matrix element\cite{Beuermann}, since
quasiparticles with momentum perpendicular to the barrier interface have the
highest probability of tunneling. For the particular case where electrons
normal to the barrier are the {\it only} ones which tunnel, there is an
exact cancellation of the one-dimensional band structure DOS by the group
velocity.

As shown by Kouznetsov {\it et al}.\cite{Kouznetsov}, the inclusion of the
group velocity and directionality to the {\it t}-{\it t}' model removes the
extraneous singularities from the calculated DOS. A varying background shape
outside the energy gap region can also be obtained by varying the tunneling
direction. Moreover, a d-wave gap symmetry may also produce a sub-gap
structure that is flat or s--wave like. The theoretical model used in this
paper develops this line of approach further with the added refinement of
using a band structure specific to optimally-doped BSCCO.

\section{THEORETICAL MODEL\protect\bigskip}

The tunneling DOS of a superconductor is determined by the imaginary part of
the single-particle Green's function,\medskip

\begin{equation}
N(E)=-\frac{1}{\pi }%
\mathop{\rm Im}%
\sum_{{\bf \ k}}|T_{{\bf k}}|^{2}G({\bf k},E)
\end{equation}

For the superconducting state,\medskip

\[
G({\bf k},E)=\frac{u_{k}^{2}}{E-E_{k}+i\Gamma }+\frac{v_{k}^{2}}{%
E+E_{k}+i\Gamma } 
\]
\medskip where $u_{k}^{2}$ and $v_{k}^{2}$ are the usual coherence factors, $%
\Gamma $ is the quasiparticle lifetime broadening factor, and $E_{k}=\sqrt{%
|\Delta ({\bf k})|^{2}+\xi _{{\bf k}}^{2}}$ with the gap function for d-wave
symmetry $\Delta ({\bf k})=\Delta _{o}[\cos (k_{x}a)-\cos (k_{y}a)]/2$ . The
tunneling matrix element $\left| T_{{\bf k}}\right| ^{2}$ is written
as\medskip

\[
\left| T_{{\bf k}}\right| ^{2}=v_{g}D({\bf k})\medskip \strut 
\]
\medskip where $v_{g}$ is the group velocity defined as $v_{g}=\left| \nabla
_{{\bf k}}\xi _{{\bf k}}\cdot {\bf n}\right| $ and {\it D}({\bf k}) is the
directionality function that has the form\cite{Ledvij},

\begin{equation}
D(k)=\exp \left[ -\frac{k^{2}-({\bf k}\cdot {\bf n})^{2}}{({\bf k}\cdot {\bf %
n})^{2}\theta _{o}^{2}}\right]
\end{equation}
The unit vector {\bf n} defines the tunneling direction, which is
perpendicular to the plane of the junction, whereas $\theta _{o}$
corresponds to the angular spread in {\it k}-space of the quasiparticle
momenta with non-negligible tunneling probability with respect to {\bf n.}
This is shown schematically in Fig. 2 for a quadrant of the Brillouin zone,
where both the Fermi surface (dashed line) and the d-wave gap are indicated.
The thick solid straight line in Fig. 2 indicates the most probable
tunneling direction defined by an angle $\theta $ with respect to the $k_{x}$
direction. The thin straight lines indicate the angular spread of tunneling
given by $\theta _{o}$. This means that for $\theta _{o}\ll 1$, only
quasiparticles with momenta extremely close to the direction of {\bf n} have
significant tunneling probability.. It should be mentioned that this
tunneling model does not take into account the possibility of an energy
dependent barrier strength\cite{Wei}, which may or may not appreciably
affect $\left| T_{{\bf k}}\right| ^{2}$\cite{Duke}. Such barrier effects
become appreciable when the tunneling particle energy is comparable to the
barrier height. The focus of our analysis is the gap region where these
effects are assumed to be small.

Another effect which is not included in the model is the zero-bias
conductance peak (ZBCP) which is sometimes observed in certain HTS
tunneling. As Fig. 1 clearly shows, there is no evidence of any ZBCP in the
SIN tunneling conductance data of BSCCO either from PCT\cite{SPIE}\cite
{Yannick} or from STM\cite{Yannick}\cite{Renner}. Recent theoretical work%
\cite{Barash1}\cite{Zhang}\cite{Widder}\cite{Tanaka}\cite{Buchholtz}\cite
{Barash2} on the effects of interface scattering by quasiparticles predicts
that a d-wave anisotropic gap should be suppressed to zero at a tunneling
interface. A redistribution of the quasiparticle spectral weight for states
in the vicinity of the interface as a result of this suppression of the gap
will result in a zero energy peak in the spectral weight. This is predicted
to result in a ZBCP in the local DOS. Furthermore, a sharp upturn in the
temperature dependence of the critical current at low temperatures is an
another consequence of the predicted ZBCP\cite{Barash2}. Some tunneling
measurements have shown the presence of zero-bias anomalies but these may be
explainable in terms of Josephson currents due to weak links\cite{Srikanth}.
Most tunneling measurements on SIN junctions do not reveal the presence of
the ZBCP with one exception being certain tunneling measurements on YBaCuO
junctions\cite{Kashiwaya}\cite{Greene}. Moreover, a survey of critical
current measurements\cite{Delin} does not show the predicted sharp upturns
in the low temperature behavior of critical currents in a wide variety of
Josephson junctions. These observations suggest that more work is required
to relate the effects of interface scattering to experimental measurements
and that the role of interface scattering and accompanying gap suppression
is unclear at present. Hence, we have not included these predicted
consequences of interface scattering in the present work.

At this point, the theoretical model will focus on the case of
optimally-doped BSCCO. We use the effective band structure extracted from
ARPES\cite{Norman},\medskip

\begin{eqnarray}
\xi _{k} &=&c_{o}+\frac{1}{2}c_{1}[\cos (k_{x}a)+\cos (k_{y}a)]+c_{2}\cos
(k_{x}a)\cos (k_{y}a)+\frac{1}{2}c_{3}[\cos (2k_{x}a)+\cos (2k_{y}a)] \\
&&\text{ +}\frac{\text{1}}{\text{2}}c_{4}[\cos (2k_{x}a)\cos (k_{y}a)+\cos
(k_{x}a)\cos (2k_{y}a)]+c_{5}\cos (2k_{x}a)\cos (2k_{y}a)  \nonumber
\end{eqnarray}
\medskip where the phenomenological parameters are $c_{o}=0.1305$, $%
c_{1}=-0.5951$, $c_{2}=0.1636$, $c_{3}=-0.0519$, $c_{4}=-0.1117$, $%
c_{5}=0.0510$. Part of the Fermi surface from this band structure is shown
as the dashed curve in Fig. 1.

\section{RESULTS}

The shape of the ``normal-state'' DOS is examined first by setting $\Delta
_{o}=0$ and imposing no directionality (similar to Fig. 1c in Ref. 15). This
case may be realized experimentally in PCT by having the tunnel current flow
radially outward from the tip into all possible {\bf k} states of the $%
Cu-O_{2}$ plane. Implied here is a subsequent c-axis transport process to
complete the current flow through the crystal. Fig. 3a shows the DOS of the
full energy band for three different $\Gamma $ values without any group
velocity effects. This, combined with no directionality, is equivalent to
setting $\left| T_{{\bf k}}\right| ^{2}=1$ in Eq. 1. The VHS is located $%
\sim $34 meV below the Fermi energy for $\Gamma $=0.003 eV and its location
shifts slightly towards higher energy as $\Gamma $ increases. The most
obvious difference between the band structure of Eq. 3 and the {\it t}-{\it t%
}' model of Ref. 15 is the absence of the second singularity at the lower
band edge. When {\it v}$_{g}$ is factored in while maintaining no
directionality, significant changes are observed (Fig. 3b). The VHSs are no
longer as distinctive as before, with peaks in the DOS having smaller
amplitudes (no rescalling of the numerical data were done in Fig. 3a and
3b). There is also a clearer shift towards higher energies in the location
of the highest peak in the DOS as $\Gamma $ increases. Additionally, the
model DOS also produces a slightly asymmetric, decreasing background at
large $|E|$. In contrast to the one-dimensional tunneling model of Harrison%
\cite{Harrison} (more appropriate for planar junctions), there is no
complete cancellation of the band structure effects by the group velocity.
However, there is clearly a strong reduction of the VHS. Note that the group
velocity is entering as a scalar quantity.

Under identical circumstances as described for Fig. 3 ( $\left| T_{k}\right|
^{2}=1$ in Eq. 1), the DOS for $\Gamma =0.003$ eV is examined with the
inclusion of the d-wave gap (Fig. 4). The DOS in Fig. 4a, which does not
include {\it v}$_{g}$, shows three distinctive peaks. The leftmost peak,
which is also the highest, is the VHS, while the remaining two are the
quasiparticle peaks. The weak ''mirror image'' of the VHS at positive energy
is a consequence of the non-zero $\Gamma $ value. As is evident in Fig. 1,
the VHS is not seen in the BSCCO data. With the inclusion of {\it v}$_{g}$,
the presence of the VHS is barely noticeable (Fig. 4b), suggesting that
group velocity effects are playing an important role in HTS tunneling. Even
though both DOS exhibit the sharp, cusp-like gap feature, there is an
observable difference in the shape of the sub-gap region. The inclusion of 
{\it v}$_{g}$ produces a tunneling DOS which has less curvature and a more
linear, V-shaped feature near zero bias.

The model DOS is compared directly with the BSCCO tunneling data of Fig. 1a,
as shown in Fig. 5. The DOS calculated with and without $v_{g}$ show a
weakly decreasing background outside the gap region. The shape inside the
gap seems to agree exceedingly well with the DOS calculated without using 
{\it v}$_{g}$ as shown in Fig. 5a. However, the extraneous presence of the
VHS in the model cannot be reconciled with the experimental data. On the
other hand, the DOS calculated with {\it v}$_{g}$ in Fig. 5b shows a better
overall agreement, especially in matching the slight asymmetry of the
quasiparticle peaks in addition to the obvious absence of the VHS. The model
in both cases produces a weakly decreasing background for $|eV|>\Delta $
that is similar to the experimental data. The cusp-like shape of the gap in
the data and the close agreement with this model suggests the possibility
that this BSCCO data is a result of a tunneling process with a very weak
directionality. This allows for quasiparticles with all momentum directions
to be involved in the tunneling process, which is a possible scenario if the
tip in the PCT is partly imbedded in the sample.

The second BSCCO data (shown in Fig. 1b and more clearly in Fig. 6) presents
features not observed in the data of Fig. 5. There are asymmetric
quasiparticle peaks, with the higher peak in the negative bias voltage
(filled states). This asymmetry is also seen in STM data on the same BSCCO
crystal\cite{Yannick}. Furthermore, the sub-gap region shows a flatter,
BCS-like feature. A smeared BCS fit of this data from Ref. 4 showed a
reasonable agreement in the sub-gap region, but could not account for the
asymmetric peaks and shape of the conductance at $|eV|>\Delta $. It is clear
that for a gap with d-wave symmetry, a flat sub-gap tunneling DOS can only
be obtained with a strong directionality effect. Also, since the voltage of
the conductance peaks is the same in Figs. 1a and 1b, it implies that the
preferred tunneling direction is near the lobe of the d-wave gap.

For this data, the full model is used, employing {\it v}$_{g}$ and the
directionality function of Eq. 2. As can be seen in Fig. 6a, there is
excellent agreement with the experimental data around the gap region up to $%
|eV|\sim 60$ meV, displaying the same degree of asymmetry and an almost flat
sub-gap structure. The value of $\Delta _{o}$ used in this model (46 meV) is
larger than that found using the smeared BCS fit (38 meV in Ref. 4 and 27.5
meV in Ref. 5). The model DOS has a large drop in the positive bias side
beyond the gap. This is due to the fact that since the directionality angle
in Eq. 2 is small with {\bf n} almost along the {\it k}$_{x}$-axis and close
to ($\pi $,0), the number of states above the Fermi surface falls off
rapidly as one approaches the edge of the Brillouin zone. As $\theta _{o}$
(which controls the width or angular spread in {\it D}({\it k})) increases
while keeping $\theta $ constant, there are not only more empty states, but
also higher energy states above the Fermi surface being summed in Eq. 1.
This explains why there are more DOS in the higher energy range in Fig. 6d.
In addition, Fig. 6 also shows that the increase in $\theta _{o}$ also
changes the shape of the sub-gap structure dramatically, approaching the
V-shaped feature found earlier. The larger $\theta _{o}$ signifies a weaker
tunneling directionality, and the V-shape arises from the inclusion of {\bf k%
} states near the line of nodes along ($\pi ,\pi $).

A natural question leading off from this is why not perform the DOS
calculation at a larger $\theta $ to include more empty and higher energy
states so as to eliminate the large DOS drop-off in the model at positive
bias? There are two major reasons why this is not done. As $\theta $
increases towards $\pi $/4, the tunneling direction approaches the node of
the d-wave gap symmetry. This means that the value of $\Delta _{o}$ in the
model has to be set at a higher, unrealistic value for this HTS compound to
be able to match the position of the quasiparticle peaks in the tunneling
data. Secondly, the smaller $\theta $ is chosen to take advantage of the
influence of the band edge on the asymmetry of the quasiparticle peaks. As
the tunneling direction approaches ($\pi ,0$), the asymmetry of the peaks is
maximized.

Hence, the value of $\theta $ that is used in Fig. 6 is not chosen
arbitrarily, but rather it is the angle which gives the appropriate degree
of asymmetry to match the experimental data. However, note that even as $%
\theta _{o}$ is increased, the degree of asymmetry of the quasiparticle
peaks of the model has not been altered as significantly as the other
features such as the decreasing asymmetry of the conductance outside the
gap. This suggests that the quasiparticle peak asymmetry is primarily a
consequence of the d-wave order parameter. Two further investigations are
done to confirm this. In Fig. 7, the same experimental BSCCO conductance
data as in Fig. 6 is compared to the model using the isotropic s-wave gap $%
\Delta $=$\Delta _{o}$. In this case, the model conductance peaks tend to be
more symmetric, and the slight asymmetry here can be attributed to the band
edge effect as shown by the normal state curve (dashed line). Secondly, we
perform the model calculation for both s- and d-wave gap symmetries at a
large $\theta $ (0.4 rad) in Eq. 2. This generates an asymmetric, normal
state background which increases towards large, positive $E$. As can be seen
in Fig. 8, even with this kind of normal state behavior, the d-wave DOS
(solid line) still shows quasiparticle peaks asymmetry that is higher in the
negative voltage side (filled states), although the degree of asymmetry is
less than that in Fig. 6. The s-wave DOS, on the other hand, still tends to
have more symmetrical quasiparticle peaks, and the minor asymmetry seems to
correspond to the shape of the underlying normal state conductance.

The origin of the asymmetry of the peaks in the tunneling DOS can be studied
further by considering the role of the tunneling matrix element $|T_{{\bf k}%
}|^{2}$ in the clean limit case ($\Gamma =0$) where

\begin{equation}
N(E)=\sum_{{\bf k}}|T_{{\bf k}}|^{2}\left[ \frac{1}{2}(1+\xi _{{\bf k}}/E_{%
{\bf k}})\delta (E_{{\bf k}}-E)+\frac{1}{2}(1-\xi _{{\bf k}}/E_{{\bf k}%
})\delta (E_{{\bf k}}+E)\right]
\end{equation}
For positive bias voltages ($E\geq 0$), the first term, involving the
coherence factor $\frac{1}{2}(1+\xi _{{\bf k}}/E_{{\bf k}})$, contributes to
the density of states because of the delta function $\delta (E_{{\bf k}}-E)$%
. In this case, the tunneling matrix element $|T_{{\bf k}}|^{2}$ selects
only a relatively short region of states in $k$-space in which the coherence
factor is greater than unity. For the majority of states integrated over (as
seen in Fig. 2) in the calculation of $N(E)$, the coherence factor is less
than unity, in fact dropping to zero at the origin for a d-wave symmetry
superconducting state.

For negative bias voltages ($E\leq 0$), the second term, involving the
coherence factor $\frac{1}{2}(1-\xi _{{\bf k}}/E_{{\bf k}})$, contributes to
the density of states because of the delta function $\delta (E_{{\bf k}}+E)$%
. In this case however, the tunneling matrix element selects out a larger
region of {\it k}-states where the coherence factor is greater than unity.
These are states below the Fermi surface in Fig. 2. The overall effect then
is to have a larger negative bias conductance compared to the positive bias
part of the conductance. This asymmetry is greater for d-wave than for
s-wave because, in the latter case, the superconducting gap does not
decrease in magnitude and become zero at ${\bf k}=0$. For the positive bias
voltage case, this tendency feature of $\Delta _{{\bf k}}=0$ at ${\bf k}=0$
enhances the reduction of the overall magnitude of the positive bias voltage 
$N(E)$ for the d-wave case, as has just been pointed out.

Therefore, the underlying asymmetry in this model is primarily due to the
d-wave gap symmetry. The proximity of the quasiparticle gap to the band edge
in Fig. 6 simply enhances the degree of asymmetry of the peaks. Also note
that in Fig. 8, there are more higher energy DOS than shown in Fig. 6 since $%
\theta $ is larger. However, as argued earlier, the value of $\Delta _{o}$
for the d-wave DOS is also considerably larger (75 meV) to be able to
produce roughly the same peak-to-peak gap value since the tunneling
direction {\bf n} in Eq. 2 is now closer to the node of the d-wave gap.

The lack of agreement between the model and the data in Fig. 6 for $V>0$
outside the gap is a significant deficiency of the model used here. There
are several possible explanations for this. The band structure for BSCCO
that was obtained from the phenomenological fit of ARPES data may be
inadequate. This is because ARPES experiments only yield reliable
information for electron emission ($V<0$) and thus only probe the filled
states of the DOS. Next is the possibility of an additional conductance
channel, such as inelastic tunneling, which has been shown to also produce
the dip feature seen in the tunneling data\cite{Kouznetsov} and an
increasing background conductance\cite{Kouznetsov}\cite{Kirtley}. There is
also the possibility of another energy band close to the top of the band
which allows for additional tunneling states.

In summary, this model presents a significant improvement over the use of a
smeared s-wave model for the DOS in analyzing tunneling conductance data in
HTS. The model makes use of a tunneling matrix element that incorporates the
group velocity of carriers, and tunneling directionality. With a band
structure which has been used previously to analyze ARPES measurement on
optimally-doped BSCCO, the model is able to duplicate the wide range of
features seen in the experimental tunneling measurements on this compound.
These features include the asymmetric quasiparticle conductance peaks in the
superconducting state, variable sub-gap conductance behavior, and the
absence of normal state band structure features, in particular the VHS. The
asymmetry of the experimentally measured quasiparticle conductance peaks,
which can be duplicated in the model using the d-wave superconducting gap,
may be a signature of the d-wave gap symmetry in BSCCO. Under no
circumstances did the model exhibit the dip features at $|eV|\sim 2\Delta $
that are consistently observed experimentally. This indicates that the dip
features arise from some other process such as strong coupling effects\cite
{Yannick} or an additional conductance channel such as inelastic tunneling%
\cite{Kouznetsov}.

The authors are grateful to Nobuaki Miyakawa for providing the experimental
data. This work was partially supported by the U.S. Department of Energy,
Division of Basic Energy Sciences - Material Sciences under contract No.
W-31-109-ENG-38, and the National Science Foundation, Office of Science and
Technology Centers under contract No. DMR 91-20000. ZY acknowledges support
from Division of Educational Programs, Argonne National Laboratory.

\newpage

\begin{center}
{\large FIGURE CAPTIONS\bigskip }
\end{center}

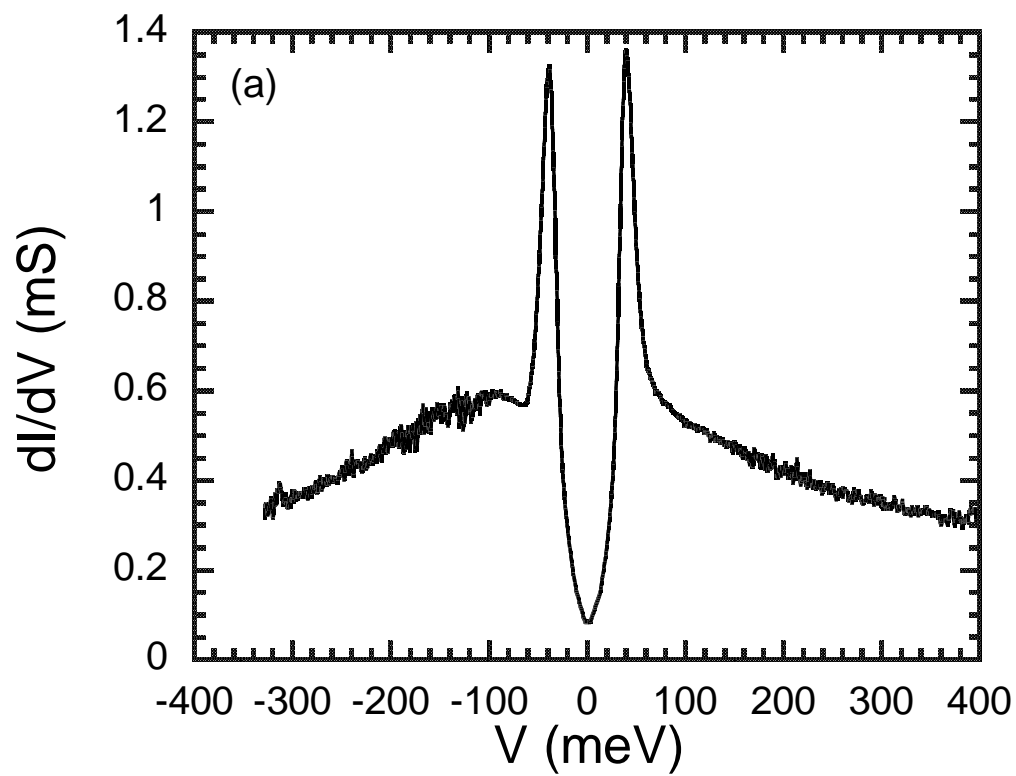
Fig. 1 : PCT SIN tunneling conductance data for optimally-doped BSCCO. Here, 
$V$ is the sample voltage with respect to the tip.\medskip

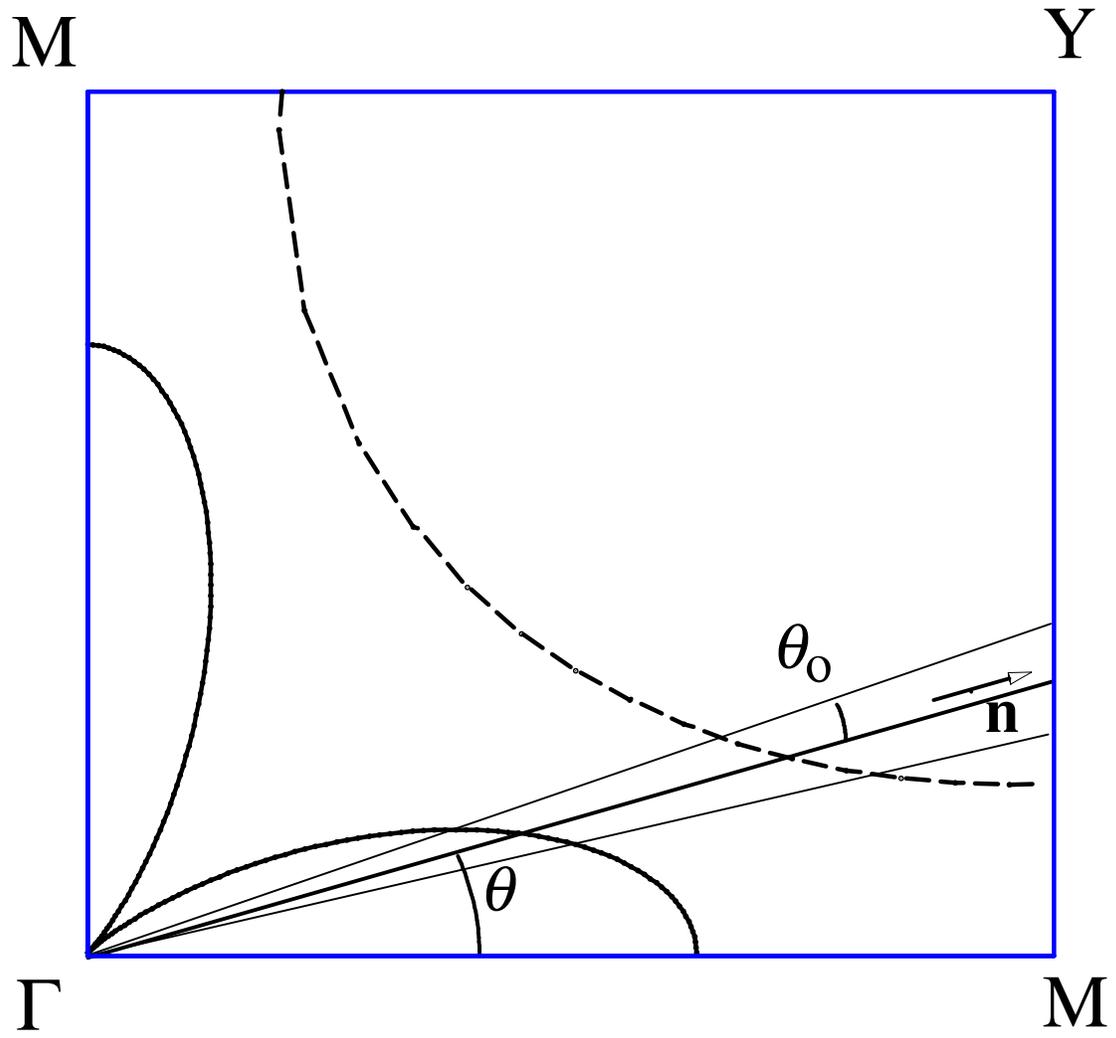
Fig. 2 : A quadrant in {\it k}-space showing the partial lobes of the d-wave
gap symmetry (solid curve), the line of directional tunneling (dark straight
line), and the angular spread $\theta _{o}$ indicated schematically by the
thin straight lines. The dashed curve represents part of the Fermi surface
specific to optimally-doped BSCCO from Ref.14.\medskip

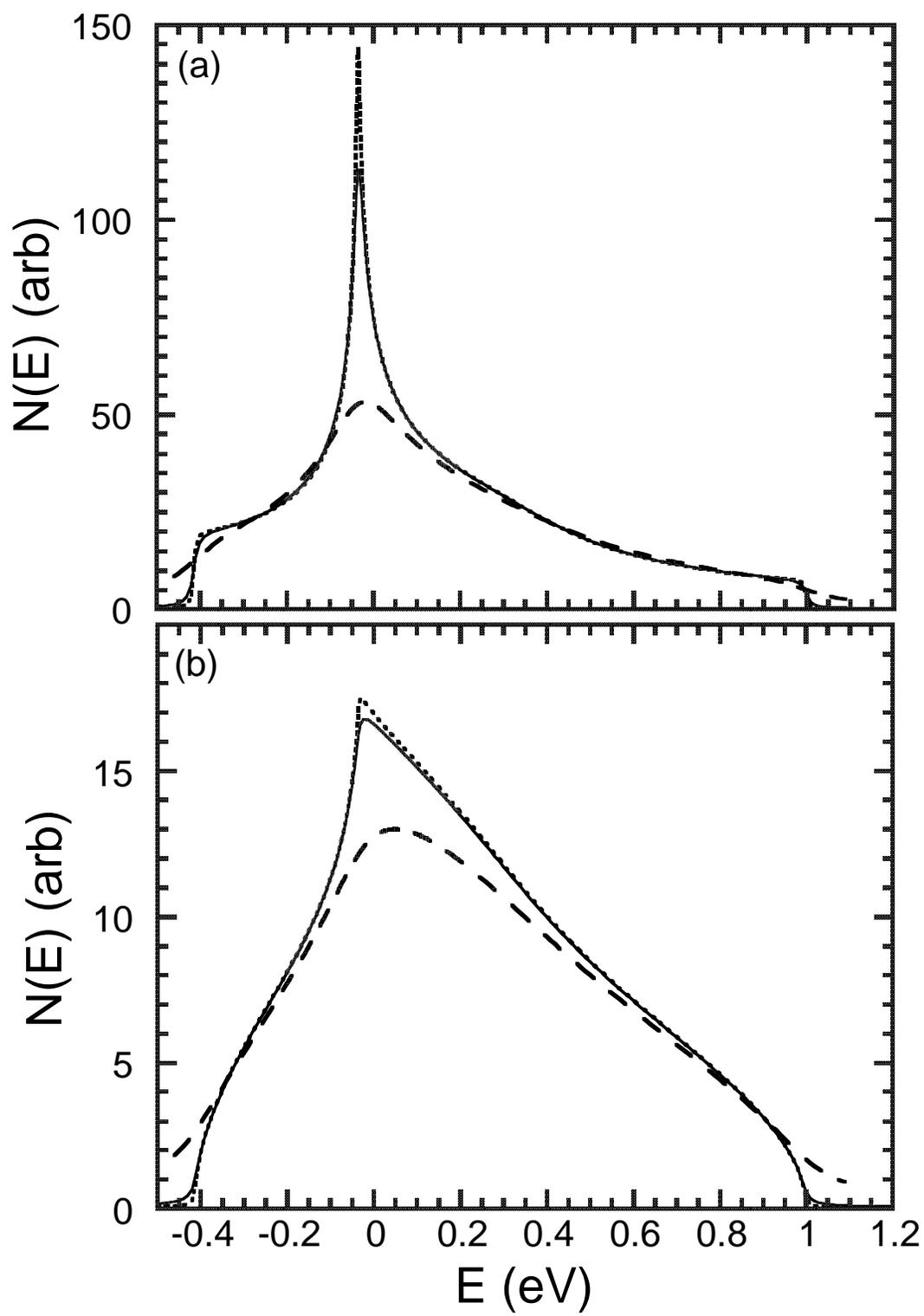
Fig. 3 : Normal state DOS calculation ($\Delta _{o}=0$) showing the full
band width for three $\Gamma $ values. The dotted line is for $\Gamma =0.003$%
eV, solid line is for $\Gamma =0.008$ eV, and the dashed line is for $\Gamma
=0.08$ eV. (a) The DOS with no {\it v}$_{g}$ and no directionality. The
single prominent peak for each curve is the van Hove singularity. (b) The
change in normal state DOS when {\it v}$_{g}$ is included.\medskip

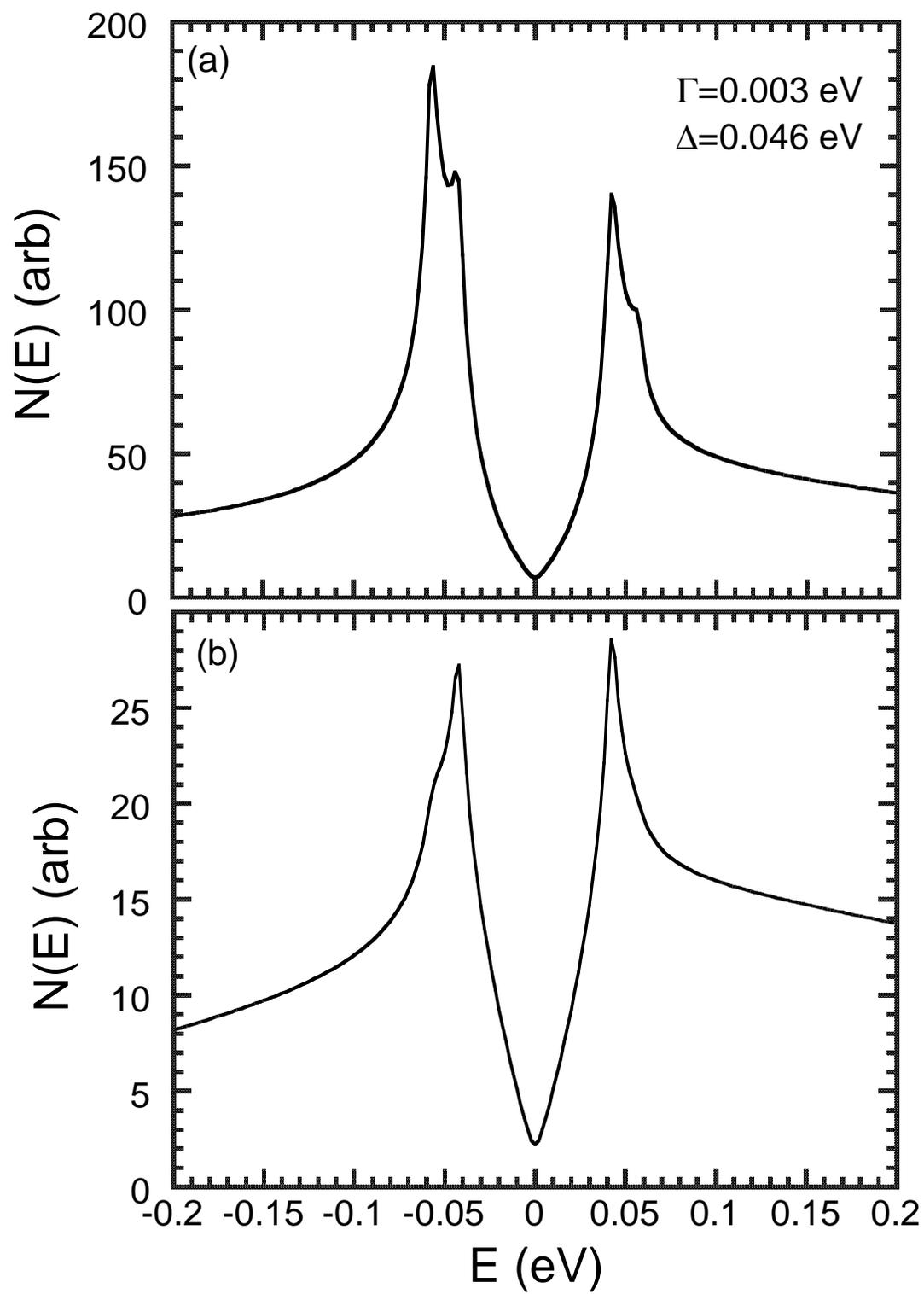
Fig. 4 : Numerical calculation of quasiparticle DOS with a d-wave gap
symmetry. (a) The DOS with no {\it v}$_{g}$ and no directionality. (b) DOS
when {\it v}$_{g}$ is included. The VHS is effectively unobservable in this
case.\medskip

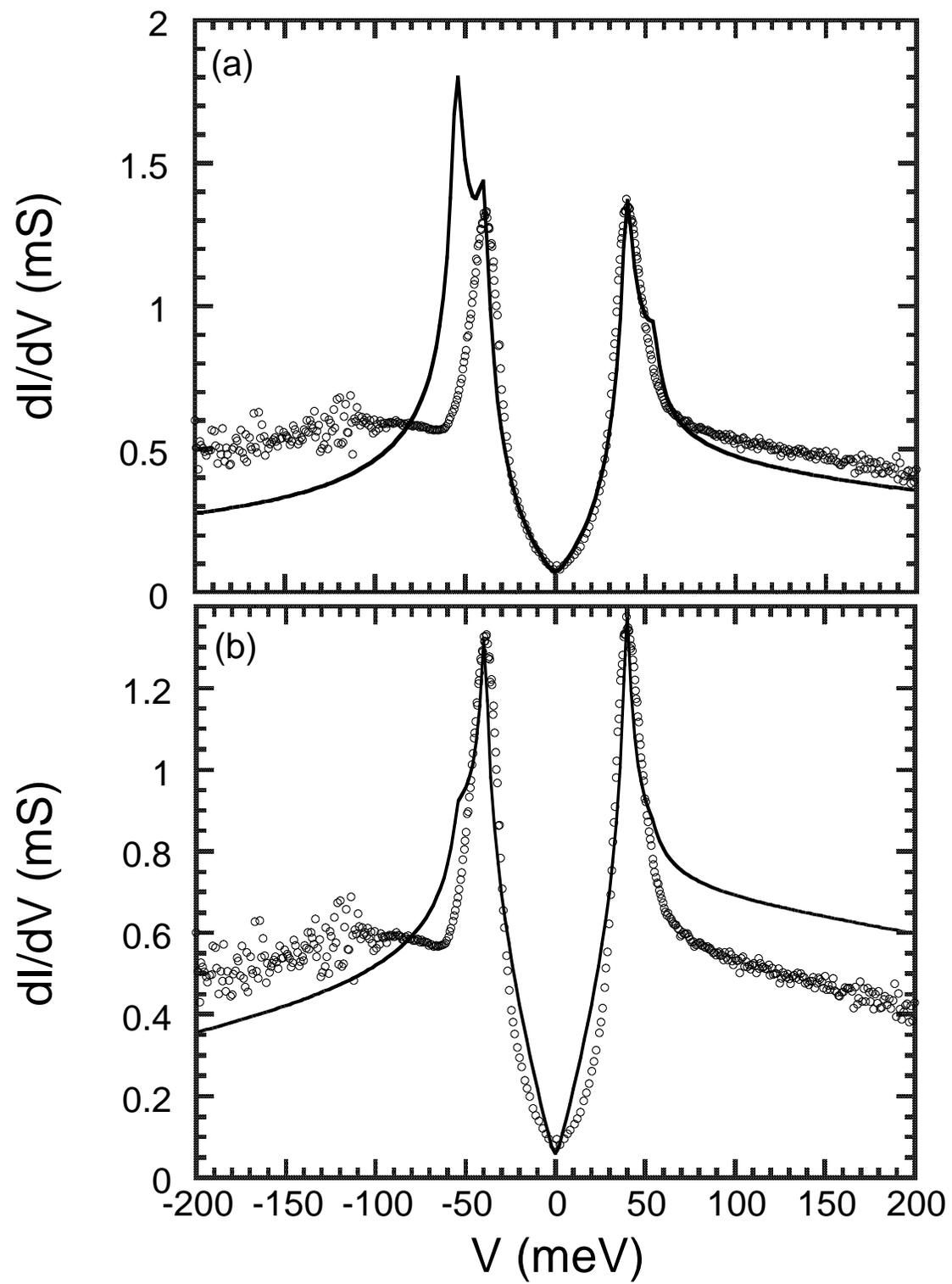
Fig. 5 : A comparison of the model (solid line) with BSCCO data from Fig. 1a
(open circles). (a) $\Delta _{o}=0.043$ eV, $\Gamma =0.002$ eV, no {\it v}$%
_{g}$, no directionality. (b) $\Delta _{o}=0.043$ eV, $\Gamma =0.001$ eV, 
{\it v}$_{g}$ present, no directionality. In each graph, the magnitude of
the model DOS was divided by a constant value to show a clearer match with
the experimental data. $\Gamma $ is different in each case to produce the
best fit to the data.\medskip

Fig. 6 : Comparison of model DOS (solid line) which includes {\it v}$_{g}$
and directionality, with BSCCO SIN tunneling conductance data from Fig. 1b
(open circles). Here $\Delta _{o}=0.046$ eV, $\Gamma =0.003$ eV, and $\theta
=0.25$ rad. for all four numerical curves. All the values of $\theta _{o}$
shown are in radians. As in Fig. 5, each of the model DOS has been rescaled
by a constant value.\medskip

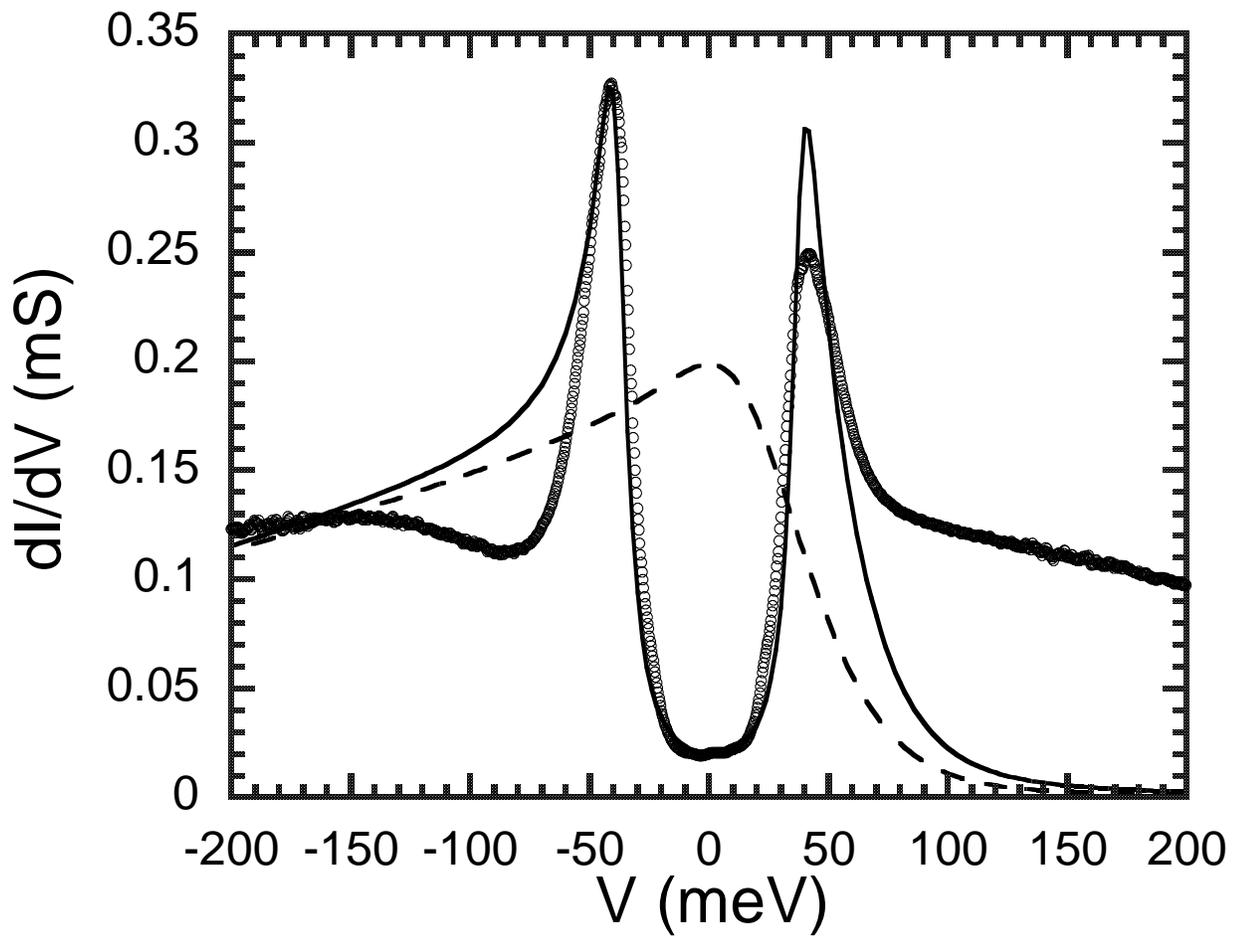
Fig. 7 : Comparison of the same BSCCO data as in Fig. 5 (open circles) with
the model using the isotropic s-wave gap symmetry $\Delta =\Delta _{o}$
(solid line). Here, $\Delta _{o}=0.038$ eV, $\Gamma =0.0035$ eV, $\theta
=0.25$ rad, $\theta _{o}$=0.05 rad to produce the best fit to the
experimental data. The dashed line is the normal state ($\Delta _{o}=0$%
).\medskip

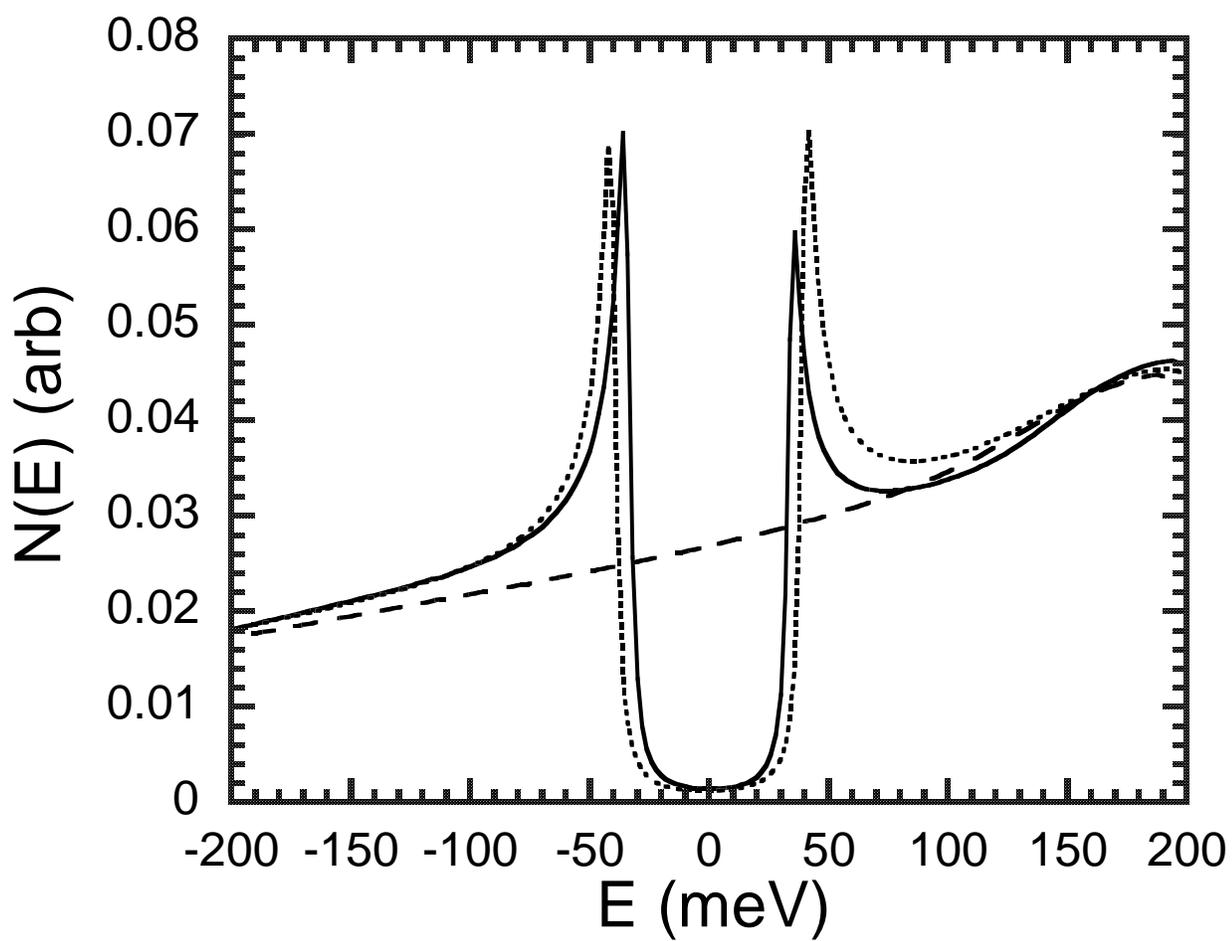
Fig. 8 : Model DOS calculations for d-wave (solid line, $\Delta _{o}=0.075$
eV), s-wave (dotted line, $\Delta _{o}=0.040$ eV), and normal state (dashed
line, $\Delta _{o}=0$). All three DOS are calculated using parameters: $%
\Gamma =0.002$ eV, $\theta =0.4$ rad, $\theta _{o}=0.02$ rad.


\begin{references}
\bibitem{tetragonal}  Here we have assumed that the crystal has tetragonal
symmetry.
\end{references}
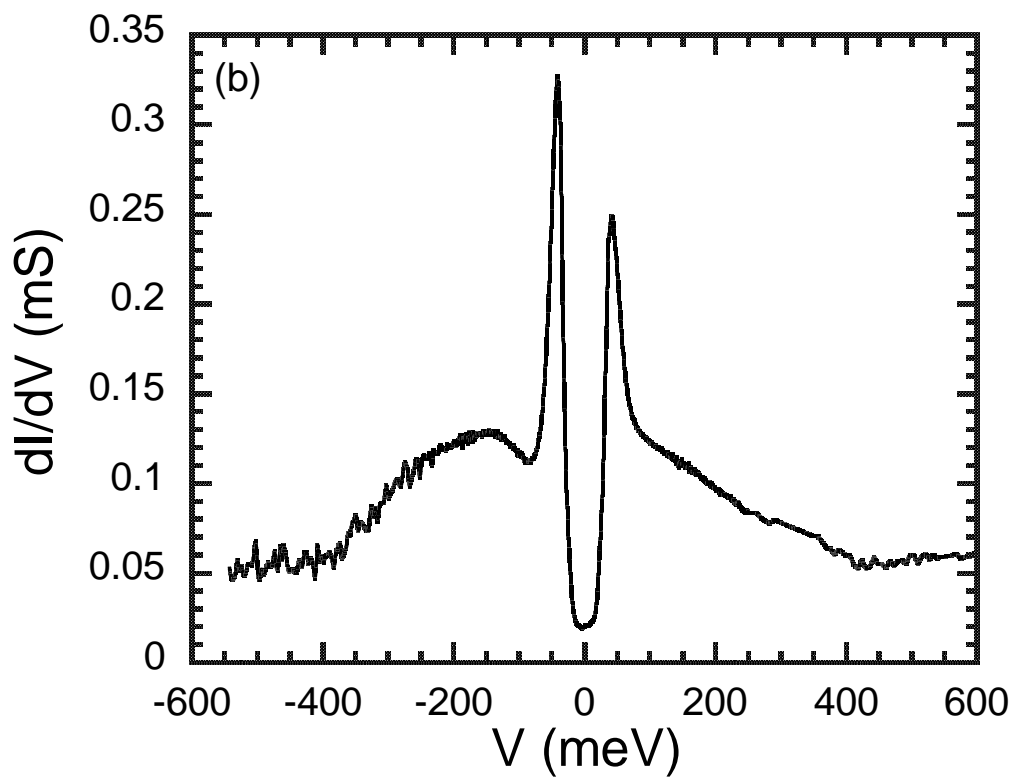
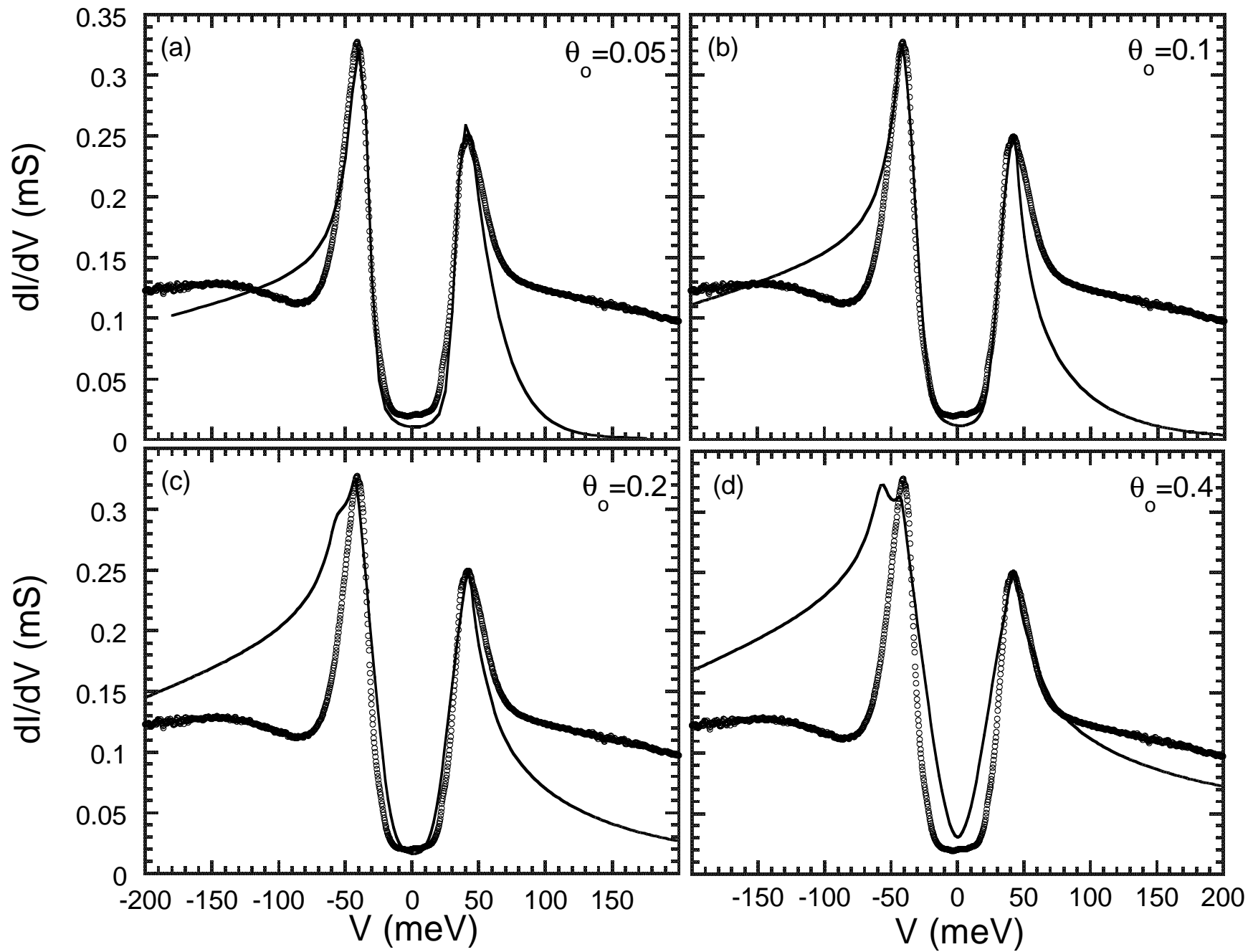
\end{document}